\documentstyle[epsfig,12pt]{article}
\newcommand{\pp}{{_{I\hspace{-0.2em}P}}}
 \begin{document}
\parskip 0.3cm
\begin{titlepage}
\begin{flushright}
CERN-TH/96-339\\
\end{flushright}

\begin{centering}
 {\large {\bf Phenomenology of  Forward Hadrons in DIS: \\ Fracture Functions   and its $Q^2$ Evolution}}\\

 \vspace{.9cm}
{\bf  D. de Florian
}
\vspace{.05in}
\\ {\it Theoretical Physics Division, CERN, CH 1211 Geneva 23, Switzerland}\\
e-mail: Daniel.de.Florian@cern.ch \\
 \vspace{.4cm}
{\bf R. Sassot} \\
\vspace{.05in}
{\it  Departamento de F\'{\i}sica, 
Universidad de Buenos Aires \\ 
Ciudad Universitaria, Pab.1 
(1428) Bs.As., Argentina} \\
e-mail: sassot@df.uba.ar \\
\vspace{1.5cm}
{ \bf Abstract}
\\ 
\bigskip
\end{centering}
 {\small   We analyse recent data on the production of forward neutrons
in deep inelastic scattering at HERA in the framework of a perturbative QCD description for semi-inclusive processes, which includes fracture functions.
Using a model estimate for the non-perturbative piece of the fragmentation process, in fairly good agreement with the available data,  we analyse the  $Q^2$ dependence of the resulting fracture functions, which is driven by  non-homogeneous evolution equations.  We also propose a measurement of the pion production cross section in the target fragmentation region  as a new test of perturbative QCD, which in this case predicts also a different $Q^2$ evolution with respect to the one given by the usual Altarelli-Parisi equations.
 }  

\vspace{0.5in}
\vfill
\begin{flushleft}
CERN-TH/96-339\\
March 1997\hfill
\end{flushleft}
\end{titlepage}
\vfill\eject

\noindent{\large \bf 1. Introduction}\\
 
During the last three decades, deep inelastic scattering (DIS) was shown to be the most efficient experiment designed to extract information about the hadron structure. With the advent of new and more powerful accelerators, it has been possible to make new tests of  QCD in a hitherto forbidden kinematical region  ($x\ll 1$ and $Q^2\gg M_p^2$), obtaining very interesting results on the behaviour of the structure function $F_2$,  in  perfect agreement with  the energy scale evolution predicted by the next to leading order (NLO) Altarelli-Parisi (AP) \cite{ap} equations and its aproximation in the double asymptotic limit \cite{ball}.

Moreover, with the analyses of the longitudinal structure function $F_L$ and the above-mentioned scaling violations, it has been possible to constrain, if not to measure, the gluon distribution over a large kinematical range and also to obtain   reliable results for the strong coupling constant $\alpha_s$. The use of evolution equations has not been limited to unpolarized scattering and its role has been preponderant in  understanding the gluonic contribution to the Ellis-Jaffe spin-dependent sum rules \cite{ellisjaffe} through the U(1) anomaly \cite{altross} and, therefore, in clarifying the situation with respect to the so-called ``spin crisis". In that way, a large  program has been developed around the measurement and the theoretical analysis of  DIS structure functions with successful results on both sides, making DIS experiments one of the  most important tests of QCD.

 Also within DIS experiments, other interesting  processes  are now under study at HERA,  specifically  events of a diffractive nature or, more precisely, with large rapidity gaps \cite{difractivo}. For  them, the proton remains almost intact after the interaction with the photon, changing only slightly its momentum and leaving the interaction region in the very forward direction without being detected. This particular kind of process has been recently 
tested by the ZEUS Collaboration \cite{zeusp}, which has included a Leading Proton Spectometer in order to detect  hadrons produced in the forward direction, and has measured the dependence of the diffractive cross section
on the transferred momentum ($t$) between the initial and the final proton. In this, they have obtained  results compatible with  previous determinations, all of them  in good agreement with the effective pomeron exchange picture,
where  the cross section is factorized  into the product of a flux of pomerons in the proton ($\propto x_\pp^{-a}$) times the, expectedly universal, distribution of partons in the pomeron. 

 More recently, the ZEUS Collaboration \cite{zeusn} has measured events where neutrons, instead of protons, are produced in the    forward direction,  obtaining a sizeable contribution of leading neutrons to the DIS cross section (as much as $10 \%$). This opens a new window to study hard processes in a new kinematical region hitherto considered  only in soft analyses.

Along these experimental improvements, a new theoretical idea has been developed in order to deal with such kind of   forward processes. In a more general perturbative picture of semi-inclusive processes, it has been shown \cite{trenvene} that both the current and the target fragmentation regions have to be considered.
The first one by means of a cross section that can be factorized into a hard cross section, parton distributions and fragmentation functions. In the second one, with another perturbatively computable hard cross section but introducing new distributions called fracture functions (measuring the probabilities of finding a parton and a hadron in the target). 

Moreover, it has also been shown that, at NLO, fracture functions are essential in order to factorize collinear singularities related to the emission of  partons in the target direction, in processes where the polarization is  either neglected  \cite{graudenz} or taken into account \cite{npb1,npb2}.

Fracture functions are expected to give the dominant contribution to cross sections for the production of leading hadrons in the target fragmentation region.   In a given kinematical range,  these distributions can be  related to the parton distributions of the object exchanged between the initial and the final state (a pomeron in $ep \rightarrow e p X$, or a $\pi^+$ in $e p \rightarrow e n X$, for example); therefore, their measurement can be used in order to obtain structure functions of ``virtual targets", which are beyond the possibilities in a ``real target" experiment. 
 It is worth noting that fracture functions allow the most appropriate way 
to factorize the hadronization properties of the forward direction into non-perturbative distributions and treat them in the context of QCD, whereas other factorization  approaches represent  effective pictures only valid in a restricted kinematical region.

In that sense, since fracture functions obey  evolution equations that differ from the Altarelli-Parisi   ones by an inhomogeneous  term  proportional to the well-known structure and fragmentation functions, their measurement  can  be taken as a test of new aspects of QCD, and at the same time can also probe the range of applicability of the  cruder factorization approaches.  Finally, since both the experimental and theoretical branches of knowledge  have been improved during the last years, it seems that a program similar to one developed for structure functions can be applied to fracture functions with, hopefully, a similar success.

Aiming in that direction, we analyse here the most recent experimental results for the production of neutrons in the forward direction in DIS, in the framework of  fracture functions,  modelling the input fracture distributions exploiting  pion exchange ideas, as it was proposed in ref. \cite{niko}, and extend that analysis to the production of pions in the target fragmentation region. This last process is expected to show
a clearly different evolution behaviour with respect to that driven by the usual evolution equations. The measurement could also be useful in order to test the non-perturbative flux proposed for pions and neutrons in protons.

 This paper is organized as follows, in the next section we briefly outline the main features of fracture functions and use the model of ref. \cite{niko},    translated into the fracture function language for the description of the fragmentation process, to analyse the data. In the third section we propose the measurement of forward  pions   and we estimate the contribution from fracture functions and  from the background composed of very forward current fragmentation events. We also analyse there the evolution of these fracture functions and discuss the possibility of finding deviations from the usual evolution. In the  last section we summarize our results and present our conclusions. \\

\noindent{\large \bf 2. Perturbative treatment of target fragmentation in the very forward region, model estimates and their comparison with the available data.}\\

In the quark-parton model, the semi-inclusive cross section for the production of a hadron $h$ from  the deep inelastic scattering of charged leptons carrying momentum $l$ off nucleons of momentum $P$, is usually described in terms of the variables  \cite{aemp}:
\begin{equation}
x=\frac{Q^2}{2 P\cdot q} , \ \ \ y=\frac{P\cdot q}{P\cdot l} ,\ \ \ 
z_h=  \frac{P \cdot h}{p \cdot q} =  \frac{E_h }{E_p (1-x)} \frac{1-\cos \theta_h}{2} \, ,
\end{equation}
where $q$ is the transferred momentum $(-q^2=Q^2)$ and $E_h$, $E_p$ and $ \theta_h$ are the produced hadron and target nucleon energies, and the angle between the hadron and the target in the centre of mass of the virtual photon-proton system, respectively.
Then, the most naive expression for the unpolarized cross section is 
\begin{eqnarray}
 \frac{d\sigma^h_p}{dx\,dy\,dz_h}  =  \frac{(1+(1-y)^2)}{2y^2}
  \sum_{i=q,\bar q} c_i\,  f_{i/p}(x)\, D_{h/i}(z_h) \, ,
\end{eqnarray}
where $c_{i}=4\pi e_{q_{i}}^2 \alpha^2/x(P+l)^2$, 
  $ f_{i/p}$ is the parton distribution of flavour $i$ and 
$D_{h/i}$ is the fragmentation function of a hadron $h$ from a parton $i$.
Next to leading order corrections to this cross section are also known, and have been shown to give a very good description of data with $\theta_h>\pi/2$ \cite{graud}, the so-called current fragmentation region.
  However, the target fragmentation region, which corresponds to $\theta_h=0$  ($z_h=0$), cannot be described with the simplified picture of eq. (2). First of all, it is easy to see that, at the lowest order, hadrons can only be produced antiparallel to the target nucleon  ($\theta_h=\pi$), excluding the forward configurations. On the other hand, going  to next to leading order, the corrections to the cross section (more precisely, the
next to leading order coefficients) develop  terms proportional to $1/z_h$. This divergence is not only related to soft emission ($E_h=0$), but also to collinear configurations where hadrons are produced in the direction of the remnant target ($\theta_h=0$). In this way, since at lowest  order hadrons cannot be produced in that direction, it is not possible to factorize the divergence, with the usual procedure, into parton distributions and fragmentation functions. 

Then, in order to  describe hadrons produced in the target fragmentation region at the lowest order, and also to be able to perform at higher orders a consistent factorization   of divergences originated in the current fragmentation region (when a parton is emitted collinearly with the target),  a new distribution has to be introduced, the so-called fracture functions, $M_{i,h/N}(x,  (1-x) z )$ \cite{trenvene,graudenz}. These distributions represent  the probability of finding a parton of flavour $i$ and a hadron $h$ in the target $N$.
It has also been shown that, within this picture, it is convenient to introduce a new variable $z = E_h/E_p (1-x)$,  which allows a discrimination between the $\theta_h=0$ and the $E_h=0$ configurations, both leading to divergences but of a different nature. The variable $z$ is equal to zero only in the case of soft hadron emission  \cite{graudenz}.

Therefore the leading order expression for the cross section becomes
\begin{eqnarray}
 \frac{d\sigma^h_p}{dx\,dy\,dz} & = & \frac{(1+(1-y)^2)}{2y^2}
  \sum_{i=q,\bar q}  c_i\,  \left[ f_{i/p}(x)\, D_{h/i}(z) \right. \nonumber \\
& +& \left.   (1-x)\, M_{i,h/p} (x, (1-x) z) \right] \, .
\end{eqnarray}
In the next to leading order, the corrections to the four cross sections that can be defined taking into account the polarizations of the initial and final state $1/2$ spin hadrons, have been computed and can be found in refs. \cite{graudenz,npb1,npb2}.

The scale dependence of fracture functions at $\cal{O}$($\alpha_s$) is driven by two kinds of processes, which contribute to the production of hadrons in the remnant target direction: the emission of collinear partons from those found in the target (the usual source of scale dependence of parton distributions, often called homogeneous evolution), and those where partons  radiated from the one to be struck by the virtual probe, fragment into the measured hadron (the so-called inhomogeneous term). These two contributions lead at leading order to the following equation:
\begin{eqnarray}
\frac{\partial}{\partial \log Q^2} M_{i,h/p}\left( \xi,\zeta, Q^2\right) = \frac{\alpha_s(Q^2)}{2\pi} \int_{\xi/(1-\zeta)}^1 \frac{du}{u} \, P^i_j (u)  \, M_{j,h/N}\left( \frac{\xi}{u},\zeta, Q^2\right) \nonumber \\
+ \frac{\alpha_s(Q^2)}{2\pi}\int^{\xi/(\xi+\zeta)}_\xi \frac{du}{\xi(1-u)} \,\hat{P}^{i,l}_j (u) \, f_{j/p} \left(\frac{\xi}{u},Q^2\right)\, D_{h/l} \left(\frac{\zeta u}{\xi (1-u)},Q^2\right) ,
\end{eqnarray}    
 where  $P^i_j (u)$ and $\hat{P}^{i,l}_j (u)$ are the regularized \cite{ap} and real \cite{veneko} Altarelli-Parisi splitting functions, respectively.

As  was mentioned in the introduction, the ZEUS Collaboration has measured DIS events identifying high-energy neutrons in the final state,  at very small angles  with respect to the proton direction ($\theta_{lab} \leq 0.75 $ mrad), in the kinematical range given by $3\times 10^{-4}< x < 6\times 10 ^{-3}$, $10< Q^2<100$ GeV$^2$ and high $x_L \equiv \tilde{E_n}/\tilde{E_p}  > 0.30$.
Here $\tilde{E_p}$ and $\tilde{E_n}$ are the target proton energy and that of the produced hadron, but in the laboratory frame, which are straightforwardly 
related to $E_h$ and $E_p$ by the appropriate boost. The variables $x_{L}$, 
$\theta_{lab}$ and $z$ are then correlated, specifically for $\theta_{lab}
\simeq 0$, $x_{L}\simeq z(1-x)$. 

The ZEUS Collaboration have reported that events with $x_L\geq 0.50$ represent a substantial fraction (of the order of  10\%) of DIS events, which means a contribution 
comparable in magnitude to the one given by the longitudinal structure function $F_L$.

In the framework of a picture for semi-inclusive processes including fracture functions, as the one outlined above, the ZEUS findings can be represented
by
 \begin{eqnarray}
 \frac{\int_{0.50}^{1-x} \frac{d\sigma^h_p}{dx\,dy\,dx_L}dx_L }  {\frac{d\sigma_p}{dx\, dy} }\equiv \frac{\int_{0.50}^{1-x}
M_2^{n/p}\left(x,\,x_L,\,Q^2  \right)dx_L }{F_2^p \left(x,Q^2\right)} \, ,
\end{eqnarray}
an expression that is exact at leading order due to the fact that, as was mentioned earlier, at this order there is no contribution from currrent fragmentation processes to the target region.  In eq. (5) we have also defined the equivalent to $F_2$ for fracture functions:
\begin{eqnarray}
M_2^{n/p}\left( x,\, x_L,\, Q^2  \right) \equiv  x \sum_i e_i^2 \,  M_{i,n/p} (x, x_L, Q^2) 
\end{eqnarray}
 and we have made explicit the integration over a finite (measured) range of $x_L$. Notice also that if the requirement $\theta_{lab}\simeq 0$ is not
fulfilled, the exact relation between $x_L$ and $z$ has to be taken into account.

In fig. 1 we show the experimental outcome for this fracture function (as defined in eq. (5) and  at $Q^2=10$ GeV$^2$, taking advantage of the negligible $Q^2$ dependence of the data), and we compare it to $F_2^p$ and $F_L^p$.
We also show the contribution to the same observable coming from  current fragmentation processes, computed taking into account the appropriate boost to the laboratory system and the cut in the $\theta_{lab}$ angle implemented in the experimental determination. It is worth noting that this last contribution is pure NLO and in fact it is about 8 orders of magnitude smaller than the experimental data\footnote{For our analysis, we use distributions from refs. \cite{grvproton,kramer}.}. Furthermore, as can be seen in the figure, even releasing the constraint on the angle for the final state neutron (but keeping the requirement of energy  fraction $>0.50$) the contribution still remains  negligible,  basically because of the kinematical suppression (the effect of the boost from the $\gamma^*p$ centre of mass system to the laboratory frame) and to the fact that this contribution begins at NLO.

Fracture functions, as  parton distributions in general, are essentially of a non-perturbative nature and have to be extracted from experiment. However, their close relation with fragmentation and structure functions allows in certain extreme cases a model estimate for them. Basically, the task 
amounts to modelling the target fragmentation process and then writing down the
result in the language of fracture functions. This procedure 
provides input fracture functions, or at least hints of their functional shape, at a definite scale, which can then be compared with actual measurements and evolved with the corresponding evolution equations. 

Recently, a very sensible model estimate for the production of forward hadrons
in DIS has been proposed \cite{niko}\footnote{In ref. \cite{joffily} an approach for the residual effects of fracture functions in the currrent fragmentation region has also been proposed.}, exploiting the idea of non-perturbative 
Fock components of the nucleon. In this approach the semi-inclusive DIS cross sections, and through them the corresponding fracture functions at a certain
input scale $Q_{0}^2$,  can be interpreted as the product of a flux of neutrons in the proton (integrated over $p_T^2$) times the structure function of the pion exchanged between them, i.e.
\begin{eqnarray}
M_2^{n/p}\left( x,\, x_L ,Q^2_0 \right) \simeq \phi_{n/p} (x_L) \, F_2^{\pi^+} \left(\frac{x}{1-x_L},Q^2_0\right) \, .
\end{eqnarray} 
Actually, ref. \cite{niko} is a proposal to measure the structure function of pions at very small $x$ using a non-perturbative computation of the flux, which is in very good agreement with experimental data on high energy neutron and $\Delta^{++}$ production in hadron-hadron collisions \cite{zoeller}. The structure functions of pions have been measured with some precision only for $x>0.10$, but there are parametrizations based on dynamical parton model assumptions (valence-like distributions at low $\mu_0^2$) \cite{grvpion}, which have been shown an impressive predictive power at small $x$ for the proton structure function \cite{grvproton},  and allow us to make a rough comparison between the model 
estimate and the data. In fig. 1 we  show the that model estimate, taking $Q^2_0$ = 10 GeV$^2$, is in  excellent agreement with the data.
In fig. 2 we also compare the $x_L$ dependence of the data and of the model prediction (normalized to the same number of events in the measured interval), showing also a good agreement. Regarding the shift in the position of the peak,
it is worth noticing that the above-mentioned ZEUS data have not been corrected either for finite acceptance effects or for those related with finite resolution, and the former may account for some, perhaps all, of the observed
shift to largers values of $x_L$  \cite{lev}. This would eventually  give an even better agreement
between the the model and the data. \\

\noindent{\large \bf 3. Scale dependence of fracture functions: pions
in the final state.}\\

The   success of the model estimate encourages us to go further and use the functional dependence of fracture functions, induced by the model and corroborated by the data,  to analyse also the $Q^2$ dependence.

We first analyse the more familar process of neutron production. Since the probability of current parton fragmentation into a neutron (given by fragmentation functions) is comparatively small with respect to that of processes originated in the target (fracture functions), no significant effects are expected in the scale evolution arising from the inhomogeneous term in this process. The evolution is mainly driven by the usual homogeneous term of the 
evolution equations leading to an almost constant ratio between the number of neutron tagged events and that of all DIS events, as observed by ZEUS.

However, the scale dependence induced in the cross section for the production of pions, at least in the kinematical region of very small $x$ and small $x_L$,
can be considerably affected by the inhomogeneity, given that soft pions are produced more copiously  from quarks than  from neutrons\footnote{We are indebted  to G. Veneziano for calling our attention to this point.}. Of course, in order to analyse these features of the evolution, an estimate for the proton to pion fracture function at some input scale $Q^2_0$ is required.  For this purpose, we can use the same ideas formerly applied to neutron production, noticing two further advantages. First, the flux to be used can be straightforwardly obtained from the one used in the last section by means of  the crossing relation \begin{equation}
\phi_{\pi^+/p} (x_L) = \phi^{\pi^+}_{n/p} (1 - x_L)\, ,
\end{equation}
where the superscript $\pi^+$ indicates that the relation is valid  only when   the contribution due to the exchange of positively charged  pions is considered. The second advantage is related to fact that   the neutron structure function is much better known than that for pions. Taking into account these features and neglecting the insignificant  contributions coming from $\Delta$  exchange, the proton to pion fracture function can be approximated by
\begin{equation}
M_2^{\pi^+/p}\left( x,\, x_L , \,Q^2_0  \right) \simeq \phi_{\pi^+/p} (x_L) \, F_2^{n} \left(\frac{x}{1-x_L}, Q^2_0\right) \, .
\end{equation}
Additionally, the direct measurement of  this fracture function could be of interest as a double test of the crude  factorization ideas.

Notice that an analogous reasoning can be made considering the exchange and production of $\pi^0$ mesons, but with  a proton instead of a neutron as final
state and intermediate particle, respectively. The study of leading $\pi^0$
production may also be possible at ZEUS \cite{lev}.

In fig. 3 we show the model estimates for proton to pion fracture functions (taking   $Q^2_0= 4$ GeV$^2$), integrated over two different bins of $x_L$, compared with the contribution coming from the current fragmentation processes. We assume here the same restrictions as in the data from the ZEUS Collaboration for neutron production\footnote{Actually, it would also be interesting to count with the $p_T$ distribution of forward hadrons in order  to establish the exact range where  fracture function contribution dominates.}. As can be seen,  higher order fragmentation  contributions are quite less significant than fracture contributions, even at the lowest $x_L$ bin.

Of course, the model is not expected to work over the whole kinematical region and, in fact,  any deviation from the scale dependence implied in eq. (9) (note that the flux is assumed to be $Q^2$ independent) would show the breakdown of the approximated factorization hypotesis. However, the ansatz in eq. (9) can be taken as an effective relation, valid at some initial value of $Q^2_0$, for which the estimated flux is adequate, and therefore provides a sensible input distribution\footnote{In fact, different choices of $Q_0^2$ do not modify the overall trends of  the evolution.}.
As usual, the correct scale dependence is that given by the evolution equations for fracture functions, and that is the aim of our next step. 

In order to study the effect of the inhomogeneity in the evolution we take different values of $x_L$, and keep them fixed while we analyse the $x$ and $Q^2$ dependence of fracture functions induced by  both the homogeneous and the
complete evolution equations.

In fig. 4 we show the result of an evolution from $Q^2_0=1$ GeV$^2$  at $x_L=0.50$. Both solutions (the homogeneous and the complete) are superimposed,   the difference being less than $0.1\%$. 
 These behaviours are perfectly compatible with the results obtained by the ZEUS Collaboration in the case of neutron production (where the inhomogeneity   contributes about 10 times less) in the same kinematical region, where no difference has been found in the evolution between $F_2$ and $M_2$.
 
However, for smaller $x_L$ the situation is completely different. As the fragmentation function increases with lower values of the argument, the inhomogeneous contribution becomes much more relevant and its effect in the evolution is sizeable.
In fact,  fig. 5 shows the evolution result for $x_L=0.10$ and $Q^2_0=1$ GeV$^2$, where the full evolution results outsize the homogeneous one by a factor of 4 at small $x$. These corrections are smaller if the ansatz of eq. (9) is assumed to be valid at values of $Q^2_0=4$ GeV$^2$ (fig. 6) and  $Q^2_0=10$ GeV$^2$ (fig. 7), but  still remain considerable and show the same 
behaviour. 

The uncertainty due to the estimate of the input fracture functions at some initial scale obviously will  be lifted as soon as they are measured;  it will be then possible to study their evolution in a model-independent way,  completely determined  by QCD, and to compare the result with the experimental data at different values of the scale. 

In fig. 8, we show for completeness the results obtained using a bin in $x_L$ between $0.10$ and $0.20$. In that case  the weight of the corrections is lowered by the inclusion of larger-$x_L$ terms, but it is noticeable that they are still about  $20 \%$, whereas corrections from the fragmentation processes to the fracture function, as in fig. 3, are less than $1\%$ in the small-$x$ range. This is due to the kinematical restrictions imposed over the final state,  but also because to the fact that the fragmentation contribution appears at the same order of $\alpha_s$ in the evolution, but one order higher in the cross section than that coming from the fracture one. \\

\noindent{\large \bf 4. Summary and conclusions}\\

In this paper, we have analysed recent experimental data on the production of
forward neutrons in DIS in terms of fracture functions, finding that the main
features of the data can be fairly reproduced by this perturbative QCD
approach, once a  non-perturbative model estimate for the input fracture
functions is given. Studying the evolution properties of these fracture functions in the specific case of forward pions in the final state, we have
found that the effects of the inhomogeneous term in the evolution equations are large and measurable, particularly in the kinematical region of very small $x$ (accessible to HERA)  and small $x_L$.
These effects are negligible for large values of $x_L$, justifying the use of the usual homogeneous Altarelli-Parisi equations for, as an example, the $t$-integrated diffractive structure function $F_2^{D(3)}( x_\pp , \beta, Q^2)$, which is just the fracture function of protons in protons   $M_2^{p/p}\left( \beta x_\pp ,\,( 1- x_\pp), Q^2  \right)$ \cite{grauven,stirling}.

The measurement of the production of pions in the target fragmentation region in DIS will be useful to check the  range of applicability of factorization ansatze  for fragmentation processes and to perform a new test of QCD by means of the analysis of  the evolution of fracture functions, which is predicted to be different from the usual one in the  kinematical region alluded to before.

\noindent{ \bf Acknowledgements}

 We gratefully acknowledge  C. Garc\'\i a Canal, G. Veneziano and G. Levman for            enlightening discussions and for carefully reading the manuscript. \\


\newpage
 
 \begin{figure}
\begin{center}
\mbox{\kern 0cm
\epsfig{file=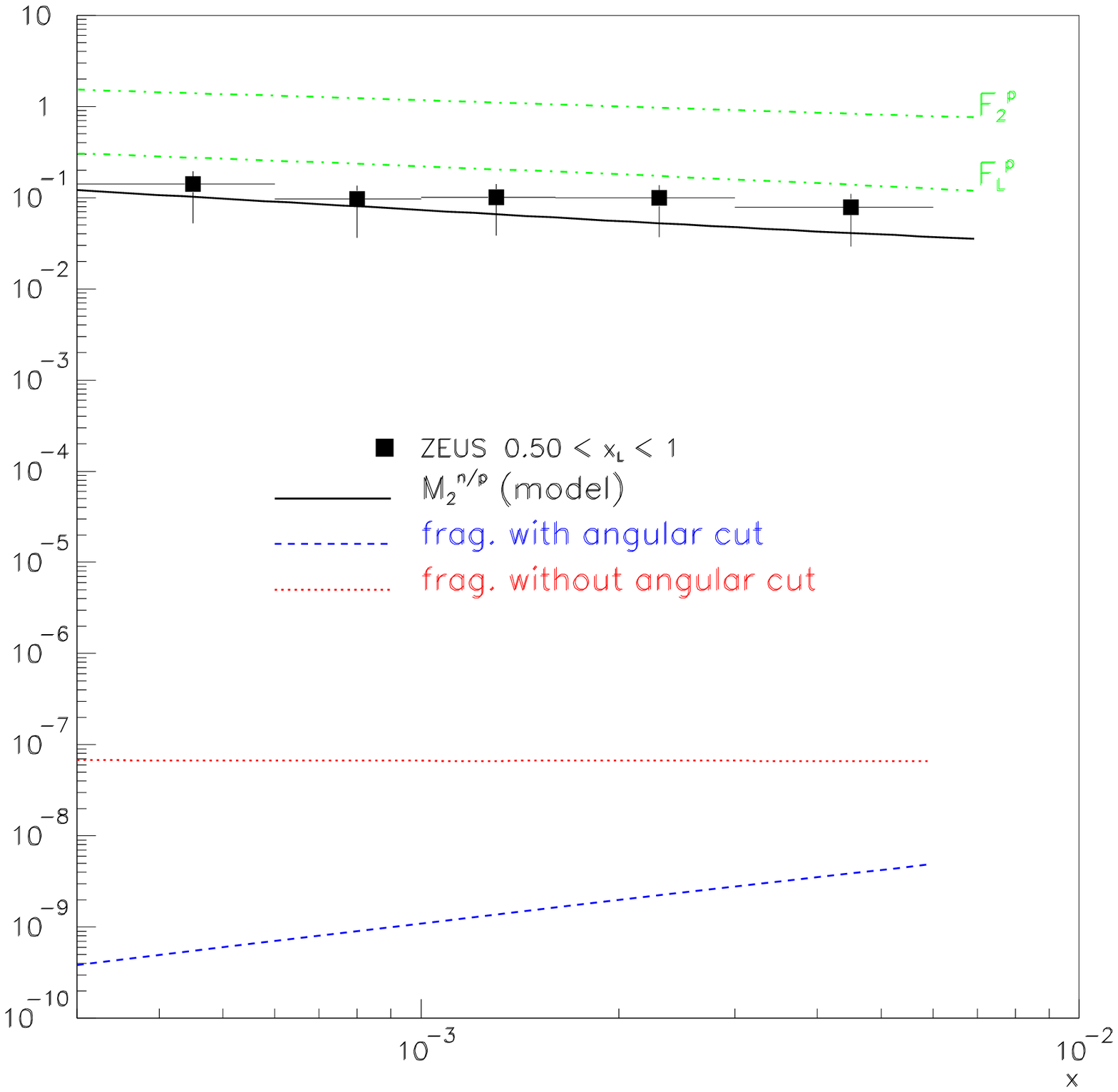,width=7.5truecm,angle=0}}
\vspace{-.7cm}
\caption{  Fracture function of neutrons in protons as measured by ZEUS compared with  the model prediction and current fragmentation contributions}
\end{center}
\end{figure}
 \begin{figure}
\begin{center}
\mbox{\kern 0cm
\epsfig{file=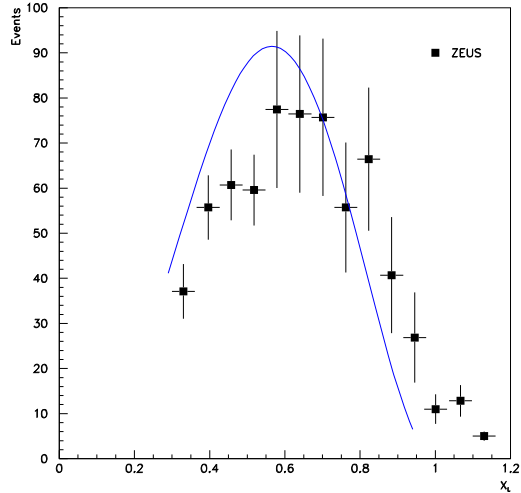,width=7.5truecm,angle=0}}
\vspace{-.7cm}
\caption{ The $x_L$ dependence of the model compared with the ZEUS data, integrated in the region $3\times 10^{-4}< x< 6\times 10^{-3}$  }
\end{center}
\end{figure}
 \begin{figure}[t]
\begin{center}
\mbox{\kern-0cm
\epsfig{file=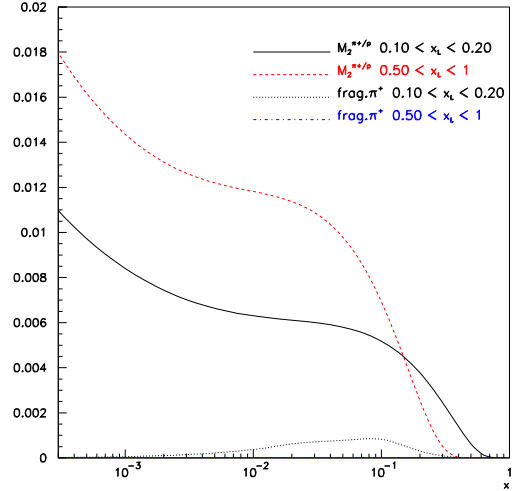,width=7.5truecm,angle=0}}
\vspace{-.7cm}
\caption{Prediction for the fracture function of $\pi^+$ in protons for two different bins of $x_L$ and the current fragmentation contribution  }
\end{center}
\end{figure}
 \begin{figure}[t]
\begin{center}
\mbox{\kern-0cm
\epsfig{file=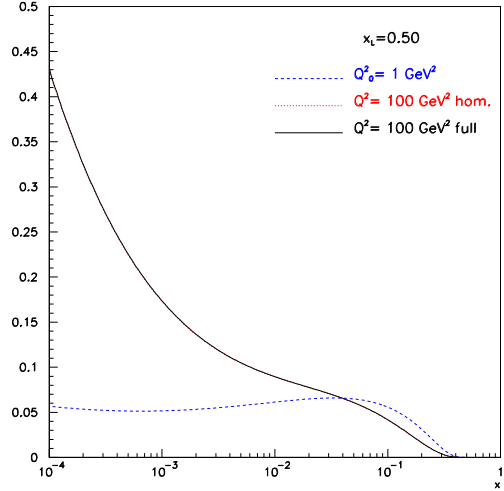,width=7.5truecm,angle=0}}
\vspace{-.7cm}
\caption{ Evolution of the fracture function of $\pi^+$ in protons for $x_L=0.50$ and $Q^2_0=1$ GeV$^2$ }
\end{center}
\end{figure}
 \begin{figure}[t]
\begin{center}
\mbox{\kern-0cm
\epsfig{file=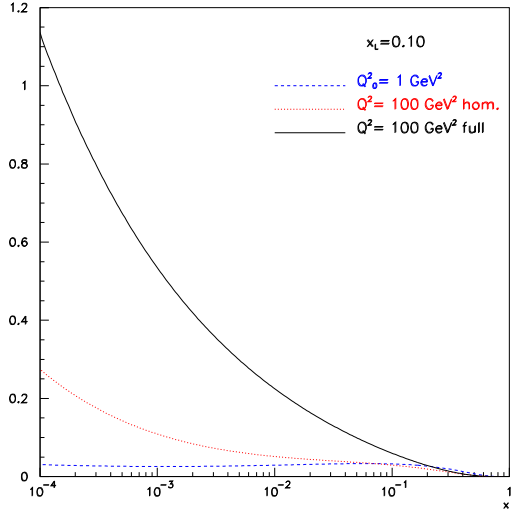,width=7.8truecm,angle=0}}
\vspace{-.5cm}
\caption{The same as in fig. 4 with $x_L=0.10$  }
\end{center}
\end{figure}
 \begin{figure}[t]
\begin{center}
\mbox{\kern-0cm
\epsfig{file=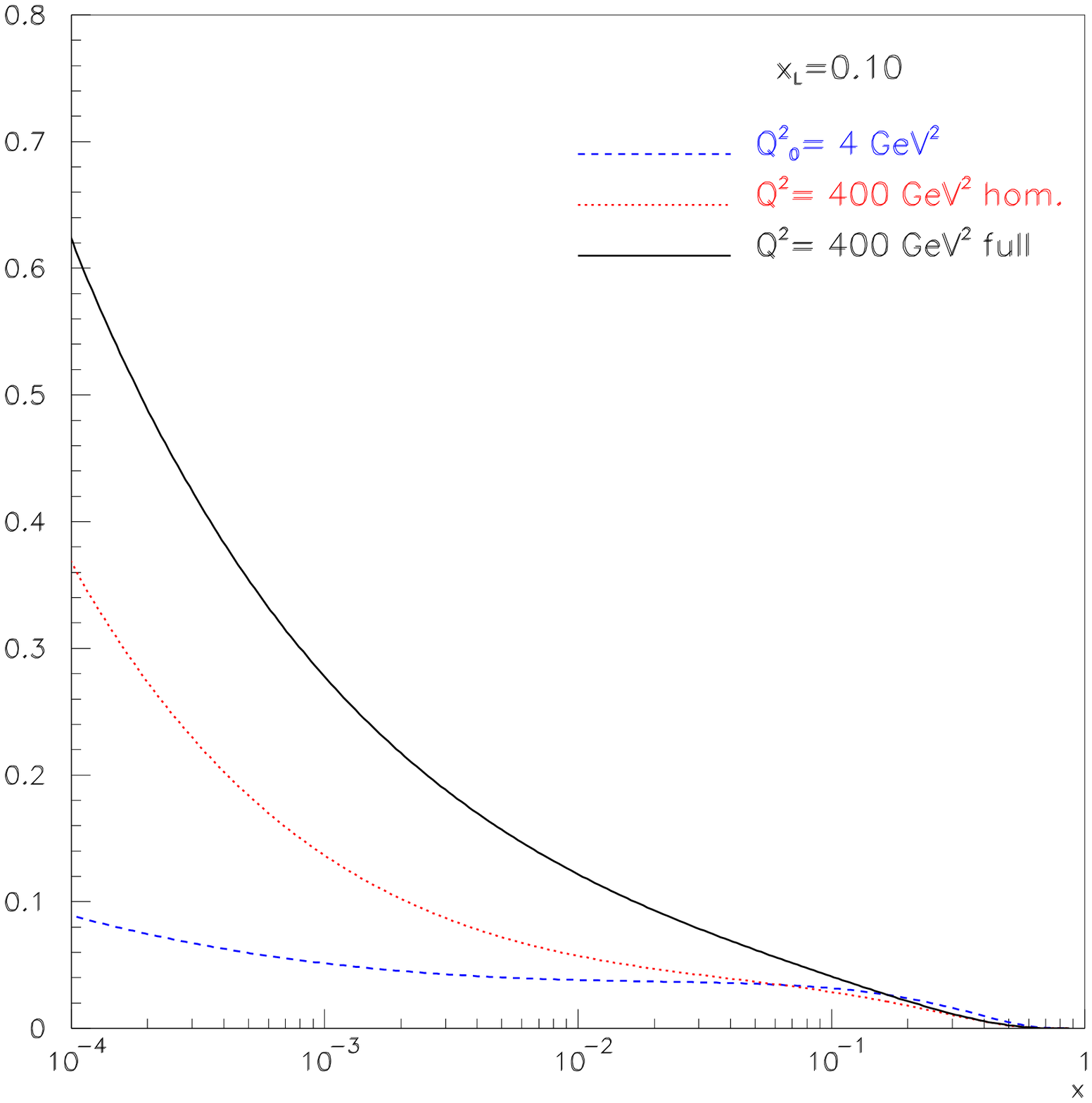,width=7.8truecm,angle=0}}
\vspace{-.5cm}
\caption{The same as in fig. 5 with $Q^2_0=4$ GeV$^2$  }
\end{center}
\end{figure}
 \begin{figure}[t]
\begin{center}
\mbox{\kern-0cm
\epsfig{file=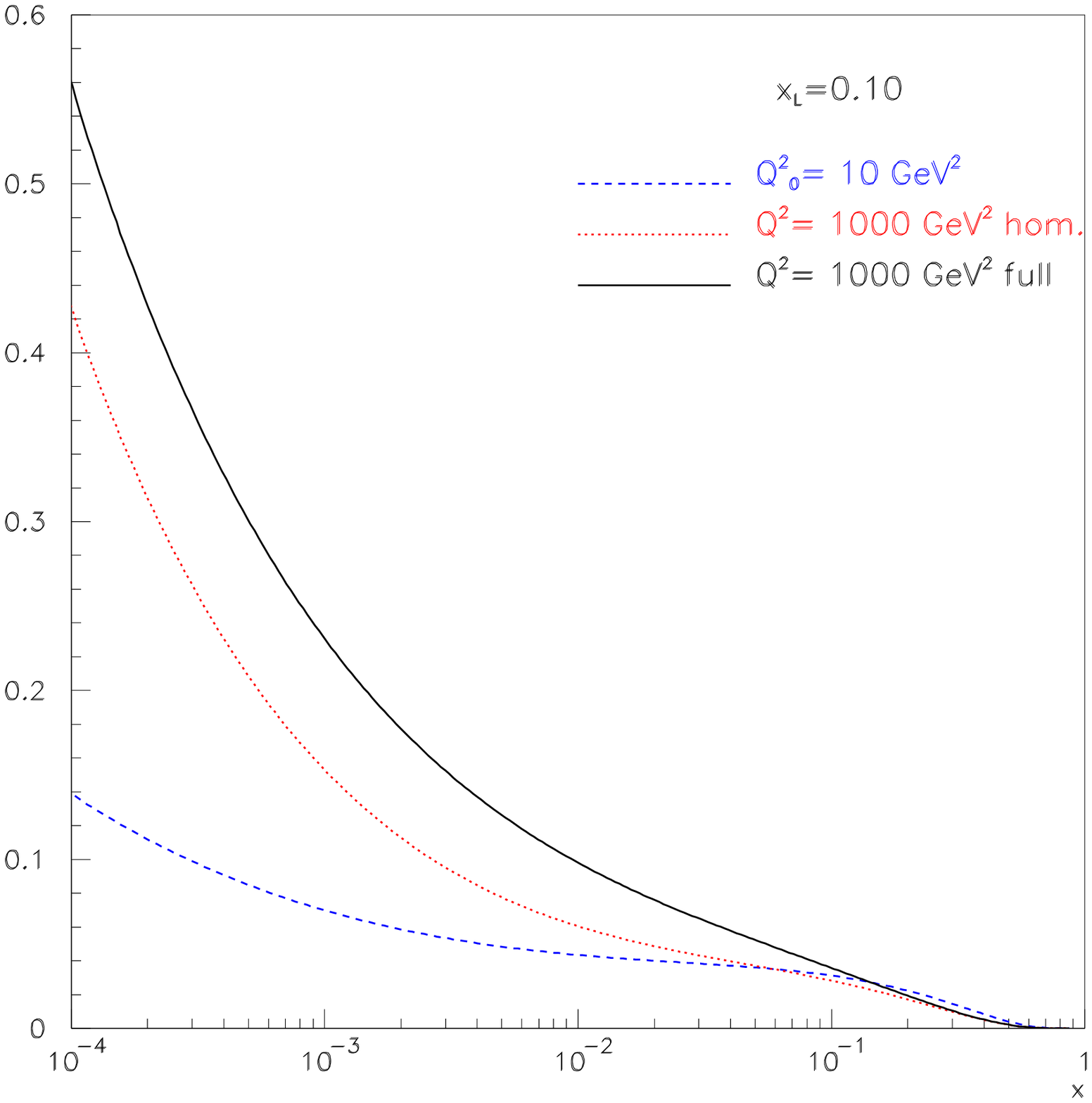,width=7.8truecm,angle=0}}
\vspace{-.5cm}
\caption{ The same as in fig. 5 with $Q^2_0=10$ GeV$^2$ }
\end{center}
\end{figure}
 \begin{figure}[t]
\begin{center}
\mbox{\kern-0cm
\epsfig{file=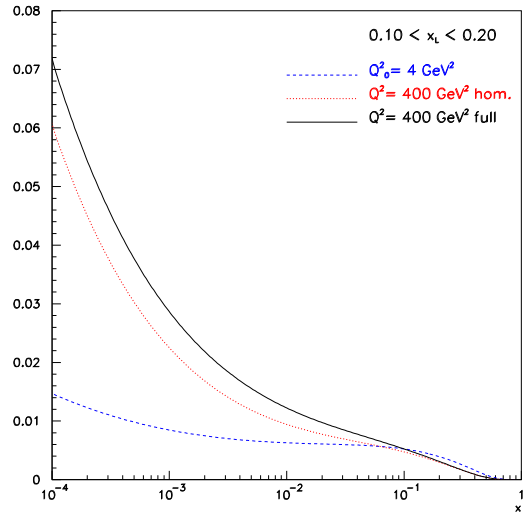,width=7.8truecm,angle=0}}
\vspace{-.5cm}
\caption{The same as in fig. 6 but integrating over a bin of $x_L$}
\end{center}
\end{figure}

\end{document}